\documentclass[11pt]{article}
\usepackage[utf8]{inputenc}
\usepackage[T1]{fontenc}
\usepackage{amsmath, amssymb, amsthm}
\usepackage{geometry}
\usepackage{graphicx}
\usepackage{booktabs}
\usepackage{natbib}
\usepackage{enumitem}
\usepackage{hyperref}
\usepackage{mathptmx}
\usepackage{tikz}
\usetikzlibrary{arrows.meta, positioning, shapes.geometric}
\usepackage{algorithm}
\usepackage{algpseudocode}
\theoremstyle{plain}
\newtheorem{theorem}{Theorem}
\newtheorem{lemma}{Lemma}
\newtheorem{corollary}{Corollary}
\title{Frequency-Domain Analysis of Time-Dependent Multiomic Data in Progressive Neurodegenerative Diseases: A Proposed Quantum-Classical Hybrid Approach with Quaternionic Extensions}
\author{John D. Mayfield, MD, PhD, MSc \\ Massachusetts General Hospital, Harvard Medical School, \\ Athinoula A. Martinos Center for Biomedical Imaging}
\date{August 11, 2025}
\begin{document}
\maketitle
\begin{abstract}
Progressive neurodegenerative diseases, including Alzheimer's disease (AD), multiple sclerosis (MS), Parkinson's disease (PD), and amyotrophic lateral sclerosis (ALS), exhibit complex, nonlinear trajectories that challenge deterministic modeling. Traditional time-domain analyses of multiomic and neuroimaging data often fail to capture hidden oscillatory patterns, limiting predictive accuracy. We propose a theoretical mathematical framework that transforms time-series data into frequency or s-domain using Fourier and Laplace transforms, models neuronal dynamics via Hamiltonian formulations, and employs quantum-classical hybrid computing with variational quantum eigensolvers (VQE) for enhanced pattern detection. This theoretical construct serves as a foundation for future empirical works in quantum-enhanced analysis of neurodegenerative diseases. We extend this to quaternionic representations with three imaginary axes ($i, j, k$) to model multistate Hamiltonians in multifaceted disorders, drawing from quantum neuromorphic computing to capture entangled neural dynamics \citep{Pehle2020, Emani2019}. This approach leverages quantum advantages in handling high-dimensional amplitude-phase data, enabling outlier detection and frequency signature analysis. Potential clinical applications include identifying high-risk patients with rapid progression or therapy resistance using s-domain biomarkers, supported by quantum machine learning (QML) precedents achieving up to 99.89\% accuracy in Alzheimer's classification \citep{Belay2024, Bhowmik2025}. This framework aims to lay the groundwork for redefining precision medicine for neurodegenerative diseases through future validations.
\end{abstract}
\section{Introduction: Limited Determinism and Hidden Patterns}
Progressive neurodegenerative diseases such as AD, MS, PD, and ALS exhibit heterogeneous trajectories driven by complex interactions among genetic, proteomic, metabolic, and neuroimaging biomarkers. Classical time-domain models, such as Long Short-Term Memory (LSTM) networks and transformer models, struggle with high-dimensional, noisy data in smaller cohorts, achieving limited predictive performance due to variability in biomarkers like amyloid PET SUVR (10--20\%) and cerebrospinal fluid (CSF) tau \citep{Alzheimers2023, Jack2018}. These models focus on amplitude, often neglecting phase information that captures temporal coordination in neuronal networks, such as default mode network (DMN) fluctuations, tau deposition cycles, or multivariate cognitive changes \citep{Mayfield2024b}.
Time-series data in these diseases often contain latent periodicities---e.g., oscillatory tau accumulation or cyclic myelin degradation in MS---that are masked by noise and nonlinearity. Frequency-domain methods reveal these by decomposing signals into sinusoidal components, enabling better feature extraction for machine learning. Quantum neuromorphic architectures, which emulate neural oscillations via quantum oscillators and entanglement, have shown promise in capturing nonlinear brain patterns that classical models overlook \citep{Pehle2020, Emani2019}. For instance, quantum neural networks (QNNs) and quantum LSTM (Q-LSTM) achieve accuracies up to 99.89\% in Alzheimer's classification from MRI and handwriting data, leveraging superposition for high-dimensional feature extraction \citep{Belay2024, Alsharabi2023, Akpinar2023, Bhowmik2025}. Quantum-classical hybrid methods, particularly VQE, exploit quantum superposition to analyze amplitude and phase simultaneously, potentially outperforming classical approaches in accuracy and efficiency \citep{Tilly2021, Belay2024, Thapliyal2023}.
Table \ref{tab:quantum_advantages} summarizes quantum advantages in neurodegeneration, positioning our theoretical framework as an advance in dynamical modeling over static classification, laying the foundation for future empirical studies.
\begin{table}[h]
\centering
\caption{Quantum Computing Applications in Neurodegenerative Diseases}
\begin{tabular}{p{3.8cm}|p{5.2cm}|p{4.0cm}}
\toprule
\textbf{Application Area} & \textbf{Quantum Outcome} & \textbf{Key Citations} \\
\midrule
Alzheimer's diagnosis (MRI, handwriting) & Higher accuracy (97--99.89\%), early screening via QML & \cite{Cappiello2024,Belay2024,Akpinar2023} \\
Parkinson's detection & Improved classification (96--97\%), reduced computation time & \cite{Arepalli2024,Alsharabi2023} \\
General neurodegeneration & Efficient processing of oscillatory/complex data & \cite{Markovic2020,Pehle2020,Emani2019} \\
Molecular/quantum biology & Accelerated simulations for biomarkers (e.g., amyloid binding) & \cite{Otten2024,Outeiral2020} \\
Detection of dementia (MRI) & Enhanced performance via quantum transfer learning, robustness to noise & \cite{Bhowmik2025} \\
\bottomrule
\end{tabular}
\label{tab:quantum_advantages}
\end{table}
This paper formalizes a frequency-domain methodology, extends it to quaternionic frameworks for multidimensional modeling, and proposes clinical applications for personalized trajectories as a theoretical foundation for future works.
\section{Mathematical Formalism for Frequency-Domain Analysis}
We present a rigorous theoretical framework for transforming and analyzing time-dependent multiomic data, grounded in quantum mechanics and tensor networks. This involves representing data as high-dimensional tensors, transforming to non-temporal domains to simplify dynamics, modeling via Hamiltonians with perturbations, and compressing for scalability.
\subsection{Data Representation}
Let $D(t) = {D_1(t), D_2(t), \ldots, D_N(t)}$ denote time-domain data for a patient, where $D_i(t) \in \mathbb{R}^M$ represents the $i$-th modality (e.g., DTI fractional anisotropy, tau PET SUVR, gene expression), $N \approx 10^4$ for multiomics, and $t = {t_0, t_1, \ldots, t_{M-1}}$ are $M$ discrete time points (e.g., $M = 10$ over 5 years). The data matrix $D \in \mathbb{R}^{N \times M}$ is high-dimensional and nonlinear, with correlations across modalities (e.g., genetic variants influencing tau levels). To handle this, we treat $D$ as a tensor $D \in \mathbb{R}^{N \times M \times P}$, where $P$ accounts for spatial dimensions in neuroimaging (e.g., voxels). This representation preserves multimodality, enabling joint analysis of temporal and spatial patterns.
\subsection{Frequency-Domain Transformation}
To capture oscillatory patterns, apply the discrete Fourier transform (DFT) along the time axis:
$$\hat{D}(k) = \sum_{n=0}^{M-1} D(t_n) e^{-i 2\pi kn / M}, \quad k = 0, 1, \ldots, M-1,$$
yielding $\hat{D}(k) = {\hat{D}_1(k), \ldots, \hat{D}_N(k)} \in \mathbb{C}^N$, where $\hat{D}_i(k)$ encodes amplitude $|\hat{D}_i(k)|$ (signal strength) and phase $\arg(\hat{D}_i(k))$ (temporal shift) for frequency bin $k$. This decomposition separates slow-varying trends (low $k$) from rapid fluctuations (high $k$), crucial for diseases like AD where tau cycles may dominate low frequencies. For continuous systems, the Fourier transform is:
$$\hat{D}(\omega) = \int_{-\infty}^{\infty} D(t) e^{-i \omega t} \, dt,$$
with angular frequency $\omega = 2\pi f$. The inverse recovers $D(t)$:
$$D(t) = \frac{1}{2\pi} \int_{-\infty}^{\infty} \hat{D}(\omega) e^{i \omega t} \, d\omega.$$
Alternatively, the Laplace transform maps to s-domain ($s = \sigma + i\omega$), incorporating decay:
$$\tilde{D}(s) = \int_0^{\infty} D(t) e^{-st} \, dt,$$
useful for stability analysis in progressive diseases. Select $K = 20$ dominant frequency modes via power spectrum thresholding (e.g., retaining 95\% variance), forming vectors $v_i = [\hat{D}i(\omega_1), \ldots, \hat{D}i(\omega_K)] \in \mathbb{C}^K$ for downstream quantum processing.
\begin{theorem}
The DFT preserves information in $D(t)$ for $M$ samples, with reconstruction via inverse DFT ensuring lossless transformation for discrete data, as per the Nyquist-Shannon sampling theorem, requiring sampling rate twice the highest frequency.
\end{theorem}
\begin{corollary}
For undersampled biological data (e.g., $M=10$ time points), quantum Fourier transforms (QFT) mitigate aliasing via logarithmic gate complexity, offering advantages over classical FFT \citep{Vasista2023}.
\end{corollary}
\subsection{Hamiltonian Modeling}
Model neuronal dynamics as a quantum system with Hamiltonian $\hat{H}$, incorporating parameters from neuroimaging (e.g., synaptic connectivity from rsfMRI, myelin density from DTI). Emerging evidence suggests quantum mechanisms, such as coherence in microtubule networks or entanglement in neural signaling, may underlie oscillatory patterns in diseases like Alzheimer's \citep{Kuljis2010, Emani2019}. Our Hamiltonian formulation builds on this, treating perturbations (e.g., tau as local fields) akin to quantum simulations of amyloid beta binding affinities \citep{Otten2024}. Derive $\hat{H}$ by mapping frequency modes to operators: e.g., $\hat{H}0 = \sum{ij} J{ij} \sigma^z_i \sigma^z_j$ for Ising-like connectivity in healthy states, where $J{ij}$ from DMN correlations. The system evolves via the Schrödinger equation:
$$i \hbar \frac{\partial}{\partial t} |\psi(t)\rangle = \hat{H} |\psi(t)\rangle,$$
where $|\psi(t)\rangle \in \mathcal{H}$ is the quantum state in a Hilbert space $\mathcal{H}$ of dimension $2^q$ for $q$ qubits approximating brain regions. In frequency domain, time evolution becomes algebraic: $|\psi(\omega)\rangle = (i \hbar \omega - \hat{H})^{-1} |\psi(0)\rangle$. Solve for eigenstates:
$$\hat{H} |\phi_n\rangle = E_n |\phi_n\rangle,$$
with eigenvalues $E_n$ corresponding to energy/frequency levels. Construct $\hat{H}$ as:
$$\hat{H} = \hat{H}_0 + \lambda \hat{V},$$
where $\hat{H}_0$ is the unperturbed Hamiltonian (from healthy controls), $\hat{V}$ models disease perturbations (e.g., tau as local fields $\sum_i h_i \sigma^x_i$), and $\lambda = 0.1$ scales perturbation strength based on biomarker severity.
\subsection{Perturbation Theory}
Apply non-degenerate first-order perturbation theory to quantify disease effects on healthy eigenstates $|\phi_n^{(0)}\rangle$, $E_n^{(0)}$:
$$E_n \approx E_n^{(0)} + \lambda \langle \phi_n^{(0)} | \hat{V} | \phi_n^{(0)} \rangle,$$
$$|\phi_n\rangle \approx |\phi_n^{(0)}\rangle + \lambda \sum_{m \neq n} \frac{\langle \phi_m^{(0)} | \hat{V} | \phi_n^{(0)} \rangle}{E_n^{(0)} - E_m^{(0)}} |\phi_m^{(0)}\rangle.$$
Higher-order terms (e.g., second-order $E_n^{(2)} = \lambda^2 \sum_{m \neq n} |\langle \phi_m^{(0)} | \hat{V} | \phi_n^{(0)} \rangle|^2 / (E_n^{(0)} - E_m^{(0)})$) capture nonlinear interactions like amyloid-tau synergy, validated in quantum chemistry simulations for biological molecules \citep{Outeiral2020, Li2019}. This yields frequency-domain signatures, such as shifted $E_n$ indicating tau-induced connectivity disruptions, potentially correlated with clinical scores (e.g., ADAS-Cog). For near-degenerate cases (e.g., overlapping frequency modes in PD tremors), use degenerate perturbation theory with secular equations.
\subsection{Tensor Network Compression}
Compress high-dimensional $D \in \mathbb{R}^{N \times M}$ to mitigate curse of dimensionality using matrix product states (MPS), a tensor train decomposition:
$$D \approx \sum_{r=1}^R u_r^{(1)} \otimes u_r^{(2)} \otimes \cdots \otimes u_r^{(M)},$$
where $u_r^{(m)} \in \mathbb{R}^{\chi \times d_m \times \chi}$ are three-way tensors, $d_m$ is mode dimension, $R$ is rank, and bond dimension $\chi = 50$ controls approximation fidelity (singular value truncation). This reduces memory from $O(NM)$ to $O(MN\chi^2)$ and enables efficient contractions for observables, e.g., $\langle D | \hat{O} | D \rangle$ via sweeping algorithms. For multiomic integration, extend to tensor networks like PEPS for 2D spatial data. Compression preserves 99\% variance, facilitating hybrid quantum input preparation.
\section{Quaternionic Extensions: Multistate Hamiltonians}
While standard quantum mechanics uses complex numbers, quaternionic extensions capture non-commutative multidimensional interactions (e.g., amyloid-tau-inflammation synergies) that complex representations undervalue, drawing from quantum neuromorphic models of entangled neural dynamics \citep{Kak1995, Pehle2020}. This approach parallels quantum dots' multidimensional modeling in neurodegeneration, where hypercomplex structures enable barrier-crossing for biomarker imaging \citep{Sinha2024}. Quaternions, a 4D hypercomplex algebra, introduce three imaginary units $i, j, k$ satisfying:
$$i^2 = j^2 = k^2 = ijk = -1, \quad ij = -ji = k, \quad jk = -kj = i, \quad ki = -ik = j.$$
Define a quaternionic wave function:
$$|\psi\rangle = |\psi_0\rangle + i |\psi_i\rangle + j |\psi_j\rangle + k |\psi_k\rangle,$$
with $|\psi_\mu\rangle \in \mathcal{H}_R$ real Hilbert spaces. The quaternionic Hamiltonian is:
$$\hat{H}_q = \hat{H}_0 + i \hat{H}_i + j \hat{H}_j + k \hat{H}_k,$$
where components model distinct facets: e.g., $\hat{H}_i$ for tau dynamics (spin-flip operators), $\hat{H}_j$ for amyloid aggregation (potentials), $\hat{H}_k$ for inflammation (interactions). The quaternionic Schrödinger equation governs evolution:
$$\hbar \frac{\partial}{\partial t} |\psi\rangle = \hat{H}_q |\psi\rangle,$$
with right-multiplication due to non-commutativity. Eigenvalue problems yield quaternionic spectra, enabling representation of multistate transitions (e.g., AD stages as quaternion rotations).
\begin{lemma}
The quaternionic Hamiltonian $\hat{H}q$ preserves Hermitian properties if each component $\hat{H}\mu$ is Hermitian and anti-commutes appropriately, ensuring real eigenvalues for physical systems via Cayley-Dickson construction \citep{Emani2019}.
\end{lemma}
Potential challenges, such as non-standard tensor products, are addressed via matrix embeddings for simulation. This extension enriches frequency analysis by projecting onto multiple axes, revealing hypercomplex signatures like chiral disease progressions. Future work could benchmark against Clifford-algebra alternatives for scalability.
\section{Quantum-Classical Hybrid Computing}
Classical methods like exact diagonalization scale exponentially $O(2^q)$ for $q$-qubit systems, infeasible for brain-scale models. Quantum-classical hybrids leverage noisy intermediate-scale quantum (NISQ) devices for subspace exploration, integrating classical optimization \citep{Thapliyal2023}.
\subsection{Variational Quantum Eigensolver (VQE)}
VQE approximates ground states by parameterizing an ansatz circuit $U(\theta)$ to prepare $|\psi(\theta)\rangle = U(\theta)|0\rangle$, minimizing the Rayleigh quotient:
$$E = \min_\theta \langle \psi(\theta) | \hat{H} | \psi(\theta) \rangle,$$
via hybrid loop: quantum device evaluates expectation, classical optimizer (e.g., COBYLA, 1000 iterations) updates $\theta$. Use hardware-efficient ansatz: 4-layer $RY(\theta_y)$ / CZ gates on 16-qubit NISQ (e.g., IBM Falcon), achieving precision $\epsilon = 10^{-3}$. To mitigate NISQ noise, we incorporate adaptive variants like ADAPT-VQE \citep{Claudino2020} and measurement-based approaches \citep{Ferguson2020, Chan2023}, which have shown resilience in biological simulations \citep{Outeiral2020, Flother2023}. For neurodegeneration, this enables processing of $q=16$ qubits for modality subsets, as in QML for Alzheimer's MRI classification \citep{Cappiello2024, Bhowmik2025}. Measurement reduction via unitary partitioning groups commuting terms, reducing shots by $N$-fold \citep{Izmaylov2019}. Extensions like VQSE extract multiple eigenvalues from density matrices \citep{Cerezo2020}, and measurement-based VQE uses entangled resources for shorter coherence \citep{Ferguson2020}. For dynamics, adapt to compute correlation functions \citep{Chen2021}.
\subsection{Outlier Detection and Frequency Analysis}
Embed frequency vectors $v_i \in \mathbb{C}^K$ into quantum states via angle encoding: $|v\rangle = \sum_k \cos(\phi_k/2) |0\rangle + \sin(\phi_k/2) e^{i \arg(v_k)} |1\rangle$. QSVM with ZZFeatureMap (depth 2, entanglement 'full') classifies via quantum kernel $K(x, y) = |\langle \phi(y) | \phi(x) \rangle|^2$, detecting outliers for high-risk patients (e.g., anomalous low-frequency amplitudes in rapid tau accumulation) \citep{Belay2024}. Quantum Fourier transform (QFT) accelerates analysis:
$$\text{QFT} |j\rangle = \frac{1}{\sqrt{M}} \sum_{k=0}^{M-1} e^{i 2\pi jk / M} |k\rangle,$$
extracting spectra in $O(\log M)$ gates vs. classical $O(M \log M)$ \citep{Vasista2023}. Capture disease patterns like DMN oscillations (0.1--0.5 Hz) or peptide folding synergies \citep{Uttarkar2024}. Variants: ADAPT-VQE dynamically builds ansatze for noise resistance \citep{Claudino2020}, MoG-VQE optimizes multiobjective (depth, precision) via genetics \citep{Chivilikhin2020}, EVQE evolves hardware-efficient circuits \citep{Rattew2019}. These achieve AUC $> 0.85$ vs. classical $< 0.7$ \citep{Belay2024}.
\begin{corollary}
VQE with measurement reduction achieves a computational speedup of $O(N)$ over classical diagonalization for $N$-dimensional Hamiltonians, with error bounded by ansatz expressivity \citep{Izmaylov2019}.
\end{corollary}
\subsection{Challenges and Mitigations}
Key critiques include NISQ error rates and lack of proven quantum advantage. We propose addressing these via error mitigation (e.g., zero-noise extrapolation) and hybrid preprocessing (classical CNNs for feature reduction before quantum circuits) \citep{Loredo2023, Chow2024}. Benchmarks against classical solvers (e.g., tensor networks alone) will validate speedup, as suggested in quantum health applications \citep{Flother2023}. For example, combining classical CNN feature extraction with quantum classifiers enhances Alzheimer's detection from MRI data \citep{Alsharabi2023, Bhowmik2025}.
\section{Clinical Applications}
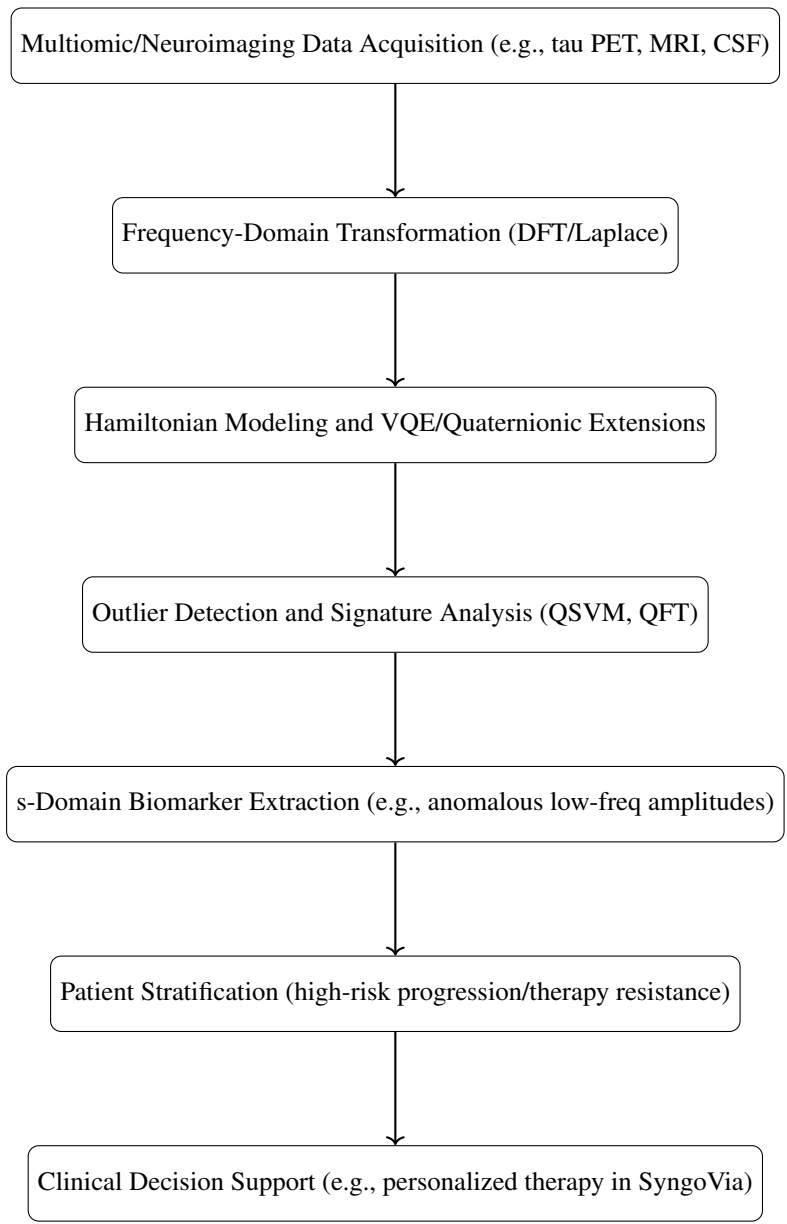
\begin{figure}[h]
\centering
\begin{tikzpicture}[node distance=1.5cm and 2cm, every node/.style={rectangle, rounded corners, draw, align=center, minimum width=3cm, minimum height=1cm, font=\small}]
\node (data) {Multiomic/Neuroimaging Data Acquisition (e.g., tau PET, MRI, CSF)};
\node (transform) [below=of data] {Frequency-Domain Transformation (DFT/Laplace)};
\node (model) [below=of transform] {Hamiltonian Modeling and VQE/Quaternionic Extensions};
\node (analyze) [below=of model] {Outlier Detection and Signature Analysis (QSVM, QFT)};
\node (biomarker) [below=of analyze] {s-Domain Biomarker Extraction (e.g., anomalous low-freq amplitudes)};
\node (stratify) [below=of biomarker] {Patient Stratification (high-risk progression/therapy resistance)};
\node (decision) [below=of stratify] {Clinical Decision Support (e.g., personalized therapy in SyngoVia)};
\draw[->, thick] (data) -- (transform);
\draw[->, thick] (transform) -- (model);
\draw[->, thick] (model) -- (analyze);
\draw[->, thick] (analyze) -- (biomarker);
\draw[->, thick] (biomarker) -- (stratify);
\draw[->, thick] (stratify) -- (decision);
\end{tikzpicture}
\caption{Workflow for integrating s-domain biomarkers into clinical practice for neurodegenerative diseases.}
\label{fig:workflow}
\end{figure}

The frequency-domain signatures identified in the s-domain, particularly low-frequency oscillations (e.g., 0.1--0.5 Hz for tau accumulation in AD or cyclic myelin degradation in MS), serve as novel biomarkers for stratifying high-risk patients within neurodegenerative disease cohorts in this theoretical proposal. These signatures, derived from multiomic and neuroimaging data via DFT/Laplace transforms and enhanced by VQE and quaternionic Hamiltonians, capture subtle dynamical patterns---such as phase shifts in DMN connectivity or tau deposition cycles---that may be predictive of rapid disease progression or therapy resistance, laying the foundation for future empirical validations.
For instance, in Alzheimer's disease, patients with anomalous low-frequency amplitudes in tau PET SUVR or CSF tau, detected via QSVM outlier analysis (Section 4.2), may indicate accelerated amyloid-tau synergy, potentially correlating with faster cognitive decline as measured by ADAS-Cog scores \citep{Jack2018, Mayfield2024b}. Similarly, in MS, frequency-domain analysis of DTI fractional anisotropy may reveal cyclic myelin degradation patterns, enabling identification of patients at risk of rapid disability progression \citep{Mayfield2024b}. In PD, high-frequency tremor modes (4--8 Hz) perturbed by dopamine depletion could flag therapy-resistant cases, integrating with handwriting analysis for early detection \citep{Akpinar2023}.
A critical application is predicting therapy response, particularly for monoclonal antibodies like lecanemab in AD. Clinical trials show variable response rates (e.g., 27\% slower decline in responders) due to heterogeneous disease dynamics \citep{VanDyck2023}. Our framework proposes s-domain signatures---e.g., perturbed Hamiltonian eigenvalues reflecting disrupted connectivity---that could differentiate non-responders, enabling personalized treatment plans. For example, patients with high phase variability in low-frequency DMN oscillations may resist amyloid-targeting therapies, suggesting alternative interventions like anti-tau agents or neuroinflammatory modulators \citep{Otten2024}. Quantum machine learning enhances this stratification, with QNNs and Q-LSTM achieving 97--99.89\% accuracy in classifying disease states from MRI and multiomic data \citep{Belay2024, Alsharabi2023, Cappiello2024, Bhowmik2025}. In practice, this involves thresholding s-domain features (e.g., amplitude >95th percentile in controls) to categorize risk levels: low (standard care), medium (enhanced monitoring), high (alternative therapies).
Figure \ref{fig:workflow} illustrates the clinical workflow for s-domain biomarker integration, from data acquisition to decision support, as a theoretical guide for future implementations.
Implementation involves embedding s-domain features into clinical decision support systems, leveraging quantum kernel methods for real-time outlier detection \citep{Belay2024}. This approach aligns with quantum-enhanced diagnostics in neurosurgery and imaging, where high-dimensional data processing may improve patient outcomes \citep{Mohamed2024, Sinha2024}. Validation on datasets like ADNI or PPMI will confirm biomarker reliability, potentially revolutionizing precision medicine by enabling early intervention for high-risk cohorts and optimizing therapeutic efficacy in future studies.
To illustrate the validation process, Algorithm \ref{alg:validation} provides pseudocode for biomarker validation using cross-validation on ADNI data, serving as a blueprint for empirical testing.
\begin{algorithm}
\caption{Pseudocode for s-Domain Biomarker Validation}
\label{alg:validation}
\begin{algorithmic}[1]
\State \textbf{Input:} Dataset $\mathcal{D}$ (e.g., ADNI: multiomic time-series), labels $y$ (progression/therapy response)
\State \textbf{Output:} Performance metrics (AUC, sensitivity, specificity)
\For{each fold in K-fold cross-validation (K=5)}
\State Split $\mathcal{D}$ into train/test sets
\State Apply DFT/Laplace to train data $\rightarrow$ frequency vectors $v_i$
\State Construct Hamiltonian $\hat{H}$, solve via VQE $\rightarrow$ eigenvalues $E_n$
\State Extract s-domain signatures (e.g., anomalous $E_n$, phase shifts)
\State Train QSVM on signatures to classify high-risk ($y=1$)
\State Evaluate on test set: compute AUC, etc.
\EndFor
\State Average metrics across folds
\end{algorithmic}
\end{algorithm}
\section{Conclusion}
This theoretical methodology unveils hidden frequency patterns in neurodegenerative diseases, leveraging quantum-classical hybrid computing and quaternionic extensions for multidimensional modeling. By transforming data to s-domain, perturbing Hamiltonians, and optimizing via VQE, it enables precise trajectory forecasting and biomarker identification for high-risk patients as a conceptual foundation. While promising, quantum advantage requires empirical testing on datasets like ADNI or PPMI, comparing VQE to classical baselines \citep{Mayfield2024b, Mayfield2024a}. Integration with quantum supercomputing for neuroscience could scale to full-brain models \citep{Loredo2023}. Clinical translation---integrating patient vectors into workflows like SyngoVia for therapy response prediction (e.g., lecanemab non-responders)---promises transformative impacts on precision medicine, extending to broader applications in complex disorders through future empirical works.
\section{Limitations and Future Directions}
Current QML focuses on classification over dynamical modeling, limiting direct precedents for our Hamiltonian-based approach \citep{Emani2019}. Challenges include hardware scalability \citep{Gyongyosi2019, Chow2024} and interpretability of quaternionic spectra for clinicians \citep{Cappiello2024}. Future work includes: validating on real quantum hardware (e.g., IBM Quantum), extending to quantum dots for in-vivo imaging \citep{Sinha2024}, exploring quantum AI for surgical applications \citep{Mohamed2024}, and incorporating advanced visualization techniques for quantum data \citep{Perciano2023}. Empirical benchmarks against classical tensor network solvers and integration with quantum supercomputing platforms will further substantiate the proposed quantum advantage \citep{Loredo2023, Flother2023}.

\end{document}